\documentstyle[psfig,preprint,aps]{revtex}

\tightenlines
\newcommand{\bea}{\begin{eqnarray}}
\newcommand{\eea}{\end{eqnarray}}
\newcommand{\be}{\begin{equation}}
\newcommand{\ee}{\end{equation}}
\newcommand{\bt}{\begin{tabular}}
\newcommand{\et}{\end{tabular}}

\newcommand{\no}{\nonumber}
\newcommand{\p}{ \mbox{\boldmath $\pi$}  }
\newcommand{\ta}{ \mbox{\boldmath $\tau$}  }
\begin{document}
\draft

\title{
            Phase Transition in the chiral $\sigma$-$\omega$ model 
            with dilatons
}
\author{\large 
         P.~Papazoglou$^{\, a}$, J.~Schaffner$^{\, b}$, 
         S.~Schramm$^{\, c}$, D. Zschiesche$^{\, a}$, H. ~St\"ocker$^{\, a}$, 
         W. ~Greiner$^{\, a}$\\[1em]}
\address{
        $^a$
        Institut f\"ur Theoretische Physik, Johann Wolfgang
        Goethe-Universit\"at,
        Postfach 111 932,\\
        D-60054 Frankfurt am Main, Germany\\[1em]
        $^b$
        Niels Bohr Institute, 
        Blegdamsvej 17, \\
        DK-2100 Copenhagen, Denmark\\[1em]
        $^c$ 
        Gesellschaft f\"ur Schwerionenforschung (GSI), Planckstra\ss{}e 1,
        Postfach 110 552,\\
        D-64220 Darmstadt, Germany}

\date{\today}

\maketitle


\begin{abstract}
We investigate the properties of different modifications to the linear $\sigma$-model (including 
a dilaton field associated with broken scale invariance) at finite 
baryon density $\rho$ and nonzero temperature $T$. The explicit 
breaking of chiral symmetry and the way the vector meson mass is generated
are significant for the appearance of a 
phase of nearly vanishing nucleon mass besides the solution describing normal nuclear matter. 
The elimination of the abnormal solution prohibits the onset of a chiral phase transition but 
allows to lower the compressibility to a reasonable range. 
The repulsive contributions from the vector mesons are responsible for the 
wide range of stability of the normal phase in the 
 ($\mu$, $T$)-plane. The abnormal solution becomes 
not only energetically preferable to the normal state at high 
temperature or density, but  also 
mechanically stable due to the inclusion of dilatons.  
\end{abstract}


\section{Introduction}
Although the underlying theory of strong interactions is believed to be known,
there is presently little hope to gain insight into the rich structure
of the nonperturbative regime at high temperature and nonzero baryon density
by solving explicitly the QCD Lagrangian. \\
Presently, theoreticians try to overcome this unsatisfactory situation
by pursuing mainly two methods: First, there is the possibility to
solve QCD numerically on a discretized space-time lattice. Reliable results are currently available only
for  finite temperature and zero baryon density.
Efforts to include dynamical fermions on the lattice are
still in their infancy and demand a huge amount of computing time.
The second possibility is to formulate an effective theory based on symmetries
which hopefully
reflects the basic features of QCD in a solvable manner. We will 
focus on the second approach since the consideration 
of symmetries and scaling may bring deep
insight into a complex problem at low computational effort \cite{kapl95}. \\
Gell-Mann and Levy \cite{gell60} succeeded early with the second kind 
of ansatz, using the linear $\sigma$-model, in order
to describe hadronic properties like pion-nucleon scattering and meson 
masses.\\
For the description of nuclear matter saturation properties it is 
necessary to introduce vector mesons so that the 
binding energy results from the cancellation of 
large repulsive and attractive contributions, in 
analogy to the phenomenologically successful 
$\sigma-\omega$-model \cite{wale74}.
Early attempts in that direction were done by Boguta who generated
the vector meson mass dynamically by   
coupling scalar fields with vector mesons in the Lagrangian \cite{bogu82}. 
Unphysical bifurcations 
could be avoided within their approach,
but one was unable to describe the chiral phase transition since the 
effective nucleon mass tended to infinity for $\rho \rightarrow \infty$. 
 The solution $m_N^{\ast}=0$ was confined to
$\rho=T=0$.  Glendenning investigated the model at high temperatures 
\cite{glen86} and found no regime at finite density and nonzero temperatures
where chiral symmetry is restored, because of the mechanical instability
of the abnormal phase. Mishustin showed that one can simultaneously avoid
bifurcations and describe a chiral phase transition at $T$=0, if one
introduces an additional field $\chi$, the dilaton, which simulates
the broken scale invariance of QCD \cite{mish93}.
By coupling dilatons to vector mesons, one is able to obtain abnormal
solutions $\sigma \simeq 0$ without making the vector field massless.
Originally, the dilaton field was introduced by Schechter in order
to mimic the trace anomaly of QCD
in an effective Lagrangian at tree level \cite{sche80}.\\
In this spirit, many authors applied chiral models to the
description of nuclear matter
properties  \cite{elli91,rodr91,heid92,cart95}.
In \cite{heid94} one was even
able to fit and describe finite nuclei as well as the widely used nonlinear
version of the Walecka-model \cite{sero86}.
This model fails to describe the
chiral phase transition, in contrast to \cite{mish93}, which exhibits
a phase transition from a normal state to an abnormal one in the sense of
Lee and Wick \cite{tdle71}.\\
The aim of the present paper is to investigate the properties of
these modified versions of the linear $\sigma$-model, which claim to give a 
satisfactory description of nuclear matter ground state properties, at 
finite temperature within the mean-field ansatz. 
In part \ref{theory} we present the model which incorporates broken 
scale and chiral symmetry. Our findings about its phase structure, the 
chiral phase transition in the ($\mu, T$)-plane, and the
temperature dependence of the nucleon effective mass are presented 
in part \ref{results}. 
\section{Theory}
\label{theory}
The linear $\sigma$-model introduced by Gell-Mann 
and Levy \cite{gell60} is 
extended to include an isoscalar vector meson $\omega$ and a scalar, isoscalar
dilaton field $\chi$ with positive parity. The scalar field $\sigma$ is the 
chiral partner of the pion and provides intermediate range 
attraction.  
The Lagrangian, which includes the particular ans\"atze  of 
\cite{mish93,elli92,heid92,heid94} reads: 
\begin{eqnarray*}
{\cal L} &=& {\cal L}_{kin}+{\cal L}_{Dirac}-{\cal V}_{vec}
-{\cal V}_{0}-{\cal V}_{CSB}
\end{eqnarray*}
\bea
\label{su2}
{\cal L}_{kin} &=& \frac{1}{2} \partial_{\mu} \sigma \partial^{\mu} \sigma
          +\frac {1}{2} \partial_{\mu} \p \partial^{\mu} \p
          +\frac {1}{2} \partial_{\mu} \chi \partial^{\mu} \chi  
          -\frac{ 1 }{ 4 } F_{ \mu \nu } F^{\mu \nu }  \\ \nonumber
{\cal L}_{Dirac} &=& \overline{N}[i\gamma_{\mu} \partial^{\mu}-g_{\omega} 
\gamma_{\mu} \omega^{\mu}
- g_{\sigma}(\sigma+i \gamma_5 \p \cdot \ta)] N \\  \nonumber
{\cal V}_{vec} &=& 
-\frac{ 1 }{ 2 } \omega_{\mu} \omega^{\mu} 
m_{\omega}^2[r(\frac{\sigma}{\sigma_0})^2
+ (1-r)(\frac{\chi}{\chi_0})^2] \\ \nonumber
{\cal V}_0 &=& -\frac{1}{2} k_0 (\frac{\chi}{\chi_0})^2 (\sigma^2+\pi^2)
+\frac{\lambda}{4}(\sigma^2+\pi^2)^2+k_1 (\frac{\chi}{\chi_0})^4 
+\frac{1}{4} \chi^4 \ln \frac{ \chi^4 }{ \chi_0^4 } 
-\frac{1}{2}\delta \chi^4 \ln \frac{\sigma^2+\p^2}{\sigma_0^2}\\ 
{\cal V}_{CSB} &=& -(\frac{\chi}{\chi_0})^2 m_{\pi}^2 f_{\pi} \sigma \nonumber \quad .
\eea
The field strength tensor reads 
$F_{\mu \nu} = \partial_{\mu} \omega_{\nu} -\partial_{\nu} \omega_{\mu}$. 
 The original $\sigma$-model is supplemented by nucleons which obey the Dirac
equation and by vector mesons whose mass is generated dynamically by the 
 $\sigma$ and $\chi$ fields. We introduce a parameter $r$ which allows 
 the vector meson 
mass to be generated by $\sigma$ and $\chi$ fields, respectively. 
The chirally invariant potential is rescaled by an appropriate power 
of the dilaton field $\chi$ in order to be scale invariant. 
The effect of the logarithmic term $ \sim \chi^4 \ln \chi$ is two-fold: 
First, it breaks scale invariance and 
leads to the proportionality $\theta_{\mu}^{\mu} \sim \chi^4$ 
as can be seen from  
\be
\theta_{\mu}^{\mu} = 4 {\cal{L}}  -\chi \frac{\partial {\cal L}}{\partial \chi}
- 2 \partial_{\mu} \chi \frac{\partial {\cal L}}
{\partial(\partial_{\mu} \chi)} = \chi^4  \quad , 
\ee
which is a consequence of the definition of scale transformations 
\cite{sche71}. Second, the logarithm leads to a non-vanishing 
vacuum expectation value for the dilaton field resulting in spontaneous 
chiral symmetry breaking. This connection comes from the term 
proportional to $\chi^2 \sigma^2$: With the breakdown 
of scale invariance the resulting mass coefficient 
becomes negative for positive $k_0$ and therefore the Nambu-Goldstone 
mode is entered. The comparison of the trace anomaly of 
QCD with that of the effective theory allows 
for the identification of the $\chi$ field with the gluon condensate: 
\be
\theta_{\mu}^{\mu} =  \langle \frac{ \beta_{QCD} }{2 g} G_{\mu \nu}^a G^{\mu \nu}_a \rangle
 \equiv (1-\delta) \chi^4
\ee
The term $\sim \delta \chi^4 \ln \sigma$ contributes to the trace anomaly 
and is motivated by the form of the 
QCD beta function 
at one loop level, for details see \cite{heid94}. 
The last term ${\cal V}_{CSB}$ breaks the chiral 
symmetry explicitly and makes the pion massive. It is scaled 
appropriately to give a dimension equal to that of the quark mass term 
$\sim m_q \overline{q} q$ of the QCD Lagrangian. \\
To investigate the phase structure of nuclear matter at finite temperature we 
adopt the mean-field approximation \cite{sero86}. 
In this approximation scheme, the fluctuations around constant vacuum 
expectation values of the 
field operators are neglected:
\bea
            \sigma(x)&=&\langle \sigma \rangle +\delta \sigma  
\rightarrow \langle \sigma \rangle \\ \no
            \chi(x)&=& \langle\chi \rangle +\delta \chi 
\rightarrow \langle \chi \rangle           \\ \no
      \omega_{\mu}(x) &=&\langle \omega \rangle \delta_{0 \mu}+ 
 \delta \omega_{\mu} 
\rightarrow  \langle \omega_0 \rangle \quad .
\eea
The fermions are treated as quantum-mechanical one-particle operators. 
The derivative terms can be neglected and only the 
time-like component of the vector meson 
$\omega \equiv \langle \omega_0 \rangle$ 
survives as we assume homogeneous and isotropic infinite nuclear
 matter. Additionally, parity conservation demands $\langle \p \rangle=0$. \\
 It is therefore straightforward to write down the thermodynamical potential  
of the grand canonical ensemble $\Omega$ per volume $V$ 
at a given temperature $T$ and chemical potential $\mu$:
\be
   \frac{\Omega}{V}= {\cal V}_{vec} + {\cal V}_0 + {\cal V}_{CSB}-{\cal V}_{vac}- 
\frac{\gamma T}{(2 \pi)^3}  
\int d^3k [\ln(1-n_k) +\ln(1-\overline{n}_k)] \quad .  
\ee
The free energy $f$ is given by 
\be
      f(\rho, T ;\sigma, \chi, \omega) = \mu \rho +\frac{\Omega}{V} \quad .
\ee 
The vacuum energy ${\cal V}_{vac}$ (the potential at $\rho=0$ and 
$T=0$) has been subtracted. $\gamma$ is the fermionic 
spin-isospin degeneracy factor (4 for the nuclear medium),
$n_k$ and $\overline{n}_k$ denote the Fermi-Dirac distribution 
functions for fermions and  
anti-fermions, respectively: 
\be
n_{k}(T, \mu^{\ast}) = 
\frac{1}{\exp \left[\left(E^{ \ast} (k)-\mu^{\ast}\right)/T\right] + 1} ;\qquad  
 \overline{n}_{k}(T, \mu^{\ast}) = 
\frac{1}{\exp \left[\left(E^{ \ast} (k)+\mu^{\ast}\right)/T\right] + 1} \quad , 
\ee
where the single particle energy is 
$E^{\ast} (k) = \sqrt{ k^2+{m_N^*}^2}$ 
with $m_N^{\ast}=g_{\sigma} \sigma$. The effective chemical potential reads
 $\mu^{\ast} = \mu-g_{\omega} \omega$. 
The meson fields are determined by extremizing 
$\frac{\Omega}{V} (\mu, T)$:  
\bea
\frac{\partial (\Omega/V)}{\partial \omega} 
&=& - \omega m_{\omega}^2 [r(\frac{\sigma}{\sigma_0})^2
+(r-1)(\frac{\chi}{\chi_0})^2] + g_{\omega} \rho = 0 \\ 
\frac{\partial (\Omega/V)}{\partial \chi} &=& 
- \omega^2 \frac{m_{\omega}^2 (1-r)}{\chi_0^2} \chi-k_0
\frac{\chi}{\chi_0^2} \sigma^2
+(4 \frac{k_1}{\chi_0^4}+1+ \ln \frac{\chi^4}{\chi_0^4}-2\delta \ln \frac{\sigma^2}{\sigma_0^2}) \chi^3
-2m_{\pi}^2 f_{\pi} \frac{\chi \sigma}{\chi_0^2} =0 \\ 
\label{su2.3}
\frac{\partial (\Omega/V)}{\partial \sigma} &=& 
-\omega^2 \frac{m_{\omega}^2 r}{\sigma_0^2}\sigma 
-k_0(\frac{\chi}{\chi_0})^2 \sigma +\lambda \sigma^3
-\delta \frac{\chi^4}{\sigma}
-m_{\pi}^2 f_{\pi}(\frac{\chi}{\chi_0})^2+g_{\sigma} \rho_s=0 \, , 
\eea 
where the scalar density is given  by
\be
\rho_s= \gamma
 \int \frac{d^3 k}{(2 \pi)^3} \frac{m_N^{\ast}}{E^{\ast}}
 (n_k+\overline{n}_k) \quad .
\ee
The vector field $\omega$ can be solved
explicitly in terms of $\sigma$ and $\chi$, yielding
\be
\omega=\frac{g_{\omega } \rho}{m_{\omega}^2(r(\frac{\sigma}{\sigma_0})^2
+ (r-1)(\frac{\chi}{\chi_0})^2)  } \qquad.
\ee
Note, that in $\omega_0$-direction the pressure is minimal, since the 
temporal and spatial components of the vector field enter with opposite 
sign and only the latter are dynamical variables.\\
In addition, one has to determine the baryon density at a 
given chemical potential via the equation 
\be
   \rho= \gamma \int \frac{d^3 k}{(2 \pi)^3}(n_k-\overline{n}_k)\quad .
\ee
The energy density and the pressure  are given by
\bea
      \epsilon &=& {\cal V}_{vec} + {\cal V}_0 + {\cal V}_{CSB}-{\cal V}_{vac} 
+\frac{\gamma}{(2 \pi)^3} \int d^3 k (E^{\ast} (k)- \mu^{\ast}) 
\stackrel{\rho=0}{\longrightarrow} 
\epsilon_{SB} =\gamma \sigma_{SB} T^4 \\ \no
p  &=& -\frac{\Omega}{V} \stackrel{\rho=0}{\longrightarrow} p_{SB}= \frac{1}{3} \gamma \sigma_{SB} T^4  
\eea
Here, the index $SB$ denotes the corresponding 
quantities in the Stefan-Boltzmann limit with $\sigma_{SB} = 7 \pi^2/120$.  
The limit $T \rightarrow 0$ can be taken straightforwardly, using
\be
    \lim_{T \rightarrow 0} T \ln (1-n_k) = E^{\ast}(k)-\mu^{\ast} \quad .
\ee
Applying the Hugenholtz-van Hove theorem \cite{hugo58}, the Fermi surface is given by 
\be
 E^{\ast}(k_F)= \sqrt{k_F^2+(g_{\sigma} \sigma)^2} = \mu^{\ast} \quad .
\ee
The scalar density and the baryon density can be determined analytically, 
yielding
\bea
 \rho_s &=&  \frac{\gamma  m_N^{\ast}}{4 \pi^2}\left[ k_F E_F^{\ast}-m_N^{\ast 2} 
\ln(\frac{k_F+E_F^{\ast}}{m_N^{\ast}})\right]\\ \nonumber
  \rho &=& \gamma \int_0^{k_F} \frac{d^3 k}{(2 \pi)^3} =
\frac{\gamma k_F^3}{6 \pi^2}    \quad .
\eea
If the dynamical vector meson mass is considered as being generated 
by $\chi$  alone and if $\delta$ is set to zero,
it is possible to solve equation \ref{su2.3} analytically:
\be
\label{su2.4}
   \chi = \chi_0 \sqrt{ \frac{\lambda \sigma^3+g_{\sigma} \rho_s}
{k_0 \sigma+m_{\pi}^2 f_{\pi}}}   \quad .
\ee
Thus, the numerical procedure is simplified to finding the root of a 
nonlinear equation of one independent variable, namely $\sigma$.  
This allows for a visualization of the phase 
structure at zero temperature. \\
In order to describe hadrons and nuclear matter 
within the model, the appropriate model parameters must be chosen.  
The pion mass is fixed at the value $m_{\pi}= 138$ MeV
which determines the parameter $k_0$ from the following relation:  
\be
 k_0 = \lambda f_{\pi}^2- \delta \frac{\chi_0^4}{f_{\pi}^2}-m_{\pi}^2 \quad .
\ee
This equation can be obtained from equation \ref{su2.3} 
by setting $\rho=T=0$ and using $\sigma_0=f_{\pi}$. In addition, one 
has to ensure that also in the vacuum 
$\frac{\partial (\Omega/V)}{\partial \chi}=0$. This leads to the 
determination of $k_1$:
\be
\label{vac}
k_1 = \frac{f_{\pi}^2}{4} (2 m_{\pi}^2+k_0-\frac{\chi_0^4}{f_{\pi}^2}) 
\quad . 
\ee
The Goldberger-Treiman relation can be used at 
the tree level to fix the coupling of the nucleons to 
the $\sigma$ field, $g_{\sigma} = \frac{m_N}{f_{\pi}}$. \\
The vector meson mass  is set to $m_{\omega} = 783$ MeV and
$\lambda$ is a free parameter which determines the $\sigma$-mass. 
The remaining two  parameters 
($g_{\omega}$ and $\chi_0$) are fitted to the ground state 
nuclear matter binding energy $E_B=\epsilon/\rho-m_N=-16$ MeV 
with zero pressure at equilibrium density $\rho_0 = 0.15$ fm$^{-3}$. 
Several parameter sets have been tested. They are listed in table 1. 
The first three rows correspond 
to the version of \cite{mish93} with $\delta=0$, which we will call hereafter the minimal model. 
There, the vector mesons are coupled only to the dilatons. 
Concerning the compressibility $K$, which should 
be around 200-400 MeV \cite{kuon95}, 
we find that a small quartic self-interaction of the $\sigma$ 
corresponding to small $\lambda$ is to be preferred in this model. 
If the logarithmic potential is included proportional to $\delta$ \cite{heid94}, 
it is possible to set $\lambda=0$ 
and therefore to lower the compressibility to reasonable values. 
Note, however, that the effective nucleon mass at $\rho_0$, which should be 
 $\approx 0.7 m_N$,  tends to increase with 
decreasing $\lambda$.   
\section{Results}
\label{results}
In order to study the properties and the impact of the different 
modifications to the minimal  
chiral model on the observables, we focus first on the phase structure 
at $T$=0 before discussing our findings at finite temperature. \\
The influence of the explicit chiral symmetry breaking term on the 
phase structure of nuclear matter is checked by computing the binding 
energy of nuclear matter 
versus the $\sigma$ field for normal nuclear density 
$\rho=\rho_0$ (Fig. 1 above) and  $\rho=4\rho_0$ (Fig. 1 below)
in the minimal version of the chiral model with $\delta=0$. 
The first and the second column correspond to the model with and without 
 explicit symmetry breaking, respectively.   
According to \cite{mish93}, the phase curve in Fig. 1a exhibits the appearance of three distinct minima: 
The first one is at $m_N^{\ast} \simeq 0.6-0.7 m_N$ (the exact value depends 
on the parametrization) which we denote as the 'normal' minimum. Besides a 
metastable minimum at roughly $m_N^{\ast} \simeq 0.2 m_N$, 
which does not play a significant role (it never 
becomes the energetically lowest state), there is a third minimum corresponding to nearly vanishing 
effective nucleon mass ($0.02 m_N$). This is the 'abnormal' minimum which becomes 
the energetically preferable state for large densities (Fig. 1b). 
There, a phase transition takes place into a chiral phase where the nucleon 
effective mass as order parameter is nearly vanishing. 
Fig. 1c shows that the exclusion of explicit symmetry breaking 
effects in the Lagrangian does change the 
phase structure even at $\rho_0$ significantly. 
Although the properties of the matter at the normal minimum are not affected, 
the exclusion of the explicit symmetry  breaking term eliminates 
the abnormal solution entirely and therefore a chiral phase transition 
does not occur. \\
There is another constraint  for the 
existence of an abnormal phase:
A  pure $\omega$-$\sigma$-coupling  without a 
dilaton admixture ($r=1$) eliminates the abnormal solution. 
This can be seen as follows: For $r=1$, an additional term 
enters the numerator of equation \ref{su2.4} 
yielding 
\be
   \chi = \chi_0 \sqrt{ \frac{\lambda \sigma^3+g_{\sigma} \rho_s 
- g_{\omega}^2 \rho^2 m_{\omega}^2 \sigma_0^2/\sigma^3}
{k_0 \sigma+m_{\pi}^2 f_{\pi}}}   \quad , 
\ee
so that $\chi$ diverges for $\sigma \rightarrow 0$. 
In fact, irrespectively of which parametrization one uses,
 $\chi$ becomes imaginary as soon as $\sigma \stackrel{<}{\sim} 0.4 m_N$. 
 No solution is possible for smaller $\sigma$ values, where an 
abnormal minimum would occur.
 We tried to lower the compressibility in
the minimal version of the chiral model presented in \cite{mish93}
and found a lower bound of $\lambda$=150 necessary 
to ensure that the abnormal state is not the energetically
lowest one at normal nuclear matter density. If an abnormal
minimum exists, at ground state density, the $\sigma^4$-term
has to contribute strongly and the compressibility cannot be lowered to 
observed values.
A way out is to permit $\delta \neq 0$ \cite{heid94}, which mimics the 
contribution of quark pairs to the QCD $\beta$-function at one loop level:
Then, it is possible to break the symmetry spontaneously even without
a quartic self-interaction, i.e. with $\lambda=0$. 
The compressibility is thus lowered to reasonable values, without
abnormal or chiral phase restoration occuring at high energy densities.\\
Let us now turn to finite temperatures. Here, the analysis gets more involved: 
three coupled equations have to be solved simultaneously. 
At low temperatures, the model exhibits a liquid-gas phase transition as 
can be seen 
from Fig. \ref{prho} (using parameter set V). 
The main difference between the minimal and the extended model sets in at high  
temperatures and densities because of the existence of the abnormal solution 
in the minimal model.  
Fig. \ref{min7} shows a contour plot of the free energy 
at $T$=170 MeV and at ground state density $\rho_0$ using set I.   
The abnormal minimum (at nearly vanishing nucleon effective 
mass) and a normal phase (at $m_N^{\ast} \simeq 0.7 m_N$) are clearly visible. 
At normal nuclear density, a chiral phase transition occurs at 
$T$=168 MeV. The phase transition is of first order, since the change 
in the free 
energy is discontinuous. \\
The calculation of the phase boundary in the
($\mu$, $T$)-plane yields surprising results if the minimal model 
is used (Fig. \ref{f300}, Set I). Along the boundary shown in the figure 
the difference between the pressure of the abnormal and normal solutions
vanishes, i.e., the transition to the chiral phase takes place. The
transition at $T$=0 was already noted in \cite{mish93}.\\
However, the extension to finite 
temperatures does not lead to a closed phase boundary, regardless which
parametrization one uses (see, e.g., triangles with $\lambda=300$, 
black circles with $\lambda=220$). The abnormal solution is stable at 
high temperatures {\it or}
at high baryon densities, but not for both. This can be seen from 
Fig. \ref{psigma}, where at four particular points in the ($\mu, T$)-plane 
of Fig. \ref{f300} the pressure as a function of the $\sigma$ field is drawn. 
The abnormal maximum of the pressure is flat (Fig. \ref{psigma}a,b) or 
it disappears completely (Fig. \ref{psigma}c) far away from the 
phase transition line. It becomes a well pronounced maximum with a 
high barrier to the normal state in the vicinity of the 
phase transition region (Fig. \ref{psigma}d).\\
The result that one has an 
open phase boundary within the plotted ($\mu$, $T$)-regime is 
unusual and counter-intuitive\footnote{However, the increase of the critical 
chemical potential at small temperatures can be shown analytically in a 
low temperature expansion \cite{tdle74}}. In contrast, in \cite{waka80} 
a closed phase boundary was obtained by 
investigating the linear $\sigma$-model including neither repulsive 
contributions from $\omega$-meson exchange nor dilatons. 
To simulate this calculation within our model we keep all 
parameters constant and change only 
the $\omega$-coupling to $g_{\omega}=6$ (black dots) 
and $g_{\omega} =0$ with varying gluon condensate (white trangles) and 
 $g_{\omega} =0$ with the gluon condensate frozen at its vacuum value 
(white circles). The presence of the 
dilaton field does not lead to the fan out of the phase 
transition curve. Nevertheless, it has the considerable effect to shift the 
transition points to roughly twice the values as compared to
 the 'non-frozen' case.  
Switching from $g_{\omega} =0$ to 
$g_{\omega}=6$ and to $g_{\omega}=8.2$, the phase boundary 
spreads out to higher densities and temperatures.
Therefore, the reason for the unusual form of the phase 
boundary is the repulsive contribution due to 
the $\omega$-meson exchange.\\
At that point we should emphasize that our results are obtained
in the framework of the mean-field approximation. The inclusion of 
quantum fluctuations in the meson fields could change our findings 
qualitatively. This wil be investigated elsewere \cite{ziesche96}.
Inclusion of resonances might lead to the closure of the boundary 
as was observed in \cite{garp79} and \cite{glen86} that taking these 
additional degrees of freedom into account, the critical densities and 
temperatures decrease. 
Another possibility to get a closed phase 
boundary might be the inclusion of a quartic self interaction for the 
vector meson, $(\omega_{\mu} \omega^{\mu})^2$, yielding 
$\omega \sim \rho^{1/3}$: the amount of repulsion at high densities 
is lowered. A detailed analysis will be found in \cite{papa96}.\\  
The extended chiral model with $\delta \neq 0$ does not show a chiral phase 
transition at all. 
The nucleon effective mass increases at high density and 
temperature\footnote{This general behavior in the chiral 
$\sigma-\omega$-model is in contrast to that suggested    
by the Nambu-Jona Lasinio model\cite{namb61,jami94} which cannot 
reproduce the binding energy of nuclear 
matter properly.}, 
as can be seen in Fig. \ref{emass}. A similar behaviour of the 
effective nucleon mass can be found for the normal phase of the minimal 
model. The difference to 
the extended model comes from the fact that -according to the phase diagram of 
figure \ref{f300}-  
a transition from high to low effective masses or vice versa can be found.\\  
In contrast to finite baryon density, almost no temperature dependence 
of the effective nucleon mass in the normal 
phase is found at $\rho=0$ until the phase transition takes place. 
In addition, the abnormal phase at $\rho=0$ 
 differs qualitatively from the one at finite density. There, 
the two fields $\sigma$ and $\chi$ vanish exactly, irrespective 
of the explicit symmetry breaking term, whereas at finite baryon 
density the $\chi$ field in the abnormal phase remains finite, as
 can be seen in Fig. \ref{min7} (there, $\chi \simeq 84$ MeV). 
When $\sigma=0$, the scalar density vanishes and from equation  
\ref{su2.4} it follows that the $\chi$ field becomes zero. Because 
the baryonic density vanishes, no singularity occurs if $\chi=0$. \\
It is also interesting to compare the high temperature phase transition 
of the Walecka-model at zero density 
studied in \cite{thei83} with that of the minimal chiral 
model (Fig. \ref{et4}). One observes  that at high temperatures the 
energy density and the 
pressure asymptotically approach  the limit of a noninteracting fermion gas. 
As in \cite{thei83}, we find that the energy density decreases with high 
temperatures whereas 
the pressure reaches its asymptotic limit from below. \\
Similar results concerning the properties of the linear $\sigma$-model at 
finite temperature were obtained in \cite{glen86}, 
which in our terminology would be the minimal model with a 
pure $\sigma-\omega$-coupling and no dilatons. 
However, there is an important difference, which results from the 
inclusion of the dilaton field $\chi$: Whereas in \cite{glen86}, 
the abnormal phase is always mechanically unstable (the pressure decreased 
with compression),  leading to the result that no region in the 
($\rho$, $T$)-plane  existed where chiral symmetry was restored, 
we find here that the abnormal or chiral restored phase is always 
mechanically stable (Fig. \ref{prho2}). 
The difference to Glendenning's work originates from the 
$\omega-\chi$- rather than $\omega-\sigma$-coupling. In contrast  
 to \cite{birs94}, where it is argued that the influence of the dilaton is 
negligible at finite density because of its high mass, we find the variation 
of the condensate $\chi$ to be essential for a mechanically stable abnormal 
phase. Similar results pointing to the importance of the dilaton field in 
nuclear matter are also obtained in \cite{kael95}, where the 
Walecka model including dilatons was studied. \\
The final question to be addressed is, whether the interesting 
(T,$\rho$)-regions can be reached in 
relativistic heavy ion collisions. For a rough estimate, we solve the 
Rankine-Hugoniot-Taub adiabate (RHTA),  which can be used as a first 
approximation for the description of nearly central collisions of fast heavy 
nuclei \cite{land75,stoe80}. The thermodynamic quantities calculated for the 
compression stage of the collision are shown in figure \ref{taub3}.
The gap in the solution of the abnormal branch comes from the disappearance of the abnormal 
maximum in the corresponding region (see, i.e, Fig. \ref{psigma}).\\
The evolution of the system in the subsequent expansion  
is calculated by the isentropes starting from a point on the Taub-adiabate 
(Fig. \ref{expansion}a). 
A minimum  at $S/A \le 2$ in the trajectory allows for the 
mechanical instability, which is suggested to cause multifragmentation. 
The expansion of the system from an abnormal initial state 
through a mixed phase into the normal state is shown in Fig. \ref{expansion}b.
Even though we cannot reach the abnormal phase with the shockfront model, 
it might be possible, i.e. with the fireball model with 
$\rho=2 \gamma_{CM} \rho_0$. 

\section{Summary and outlook}
The properties of the linear 
$\sigma$-model presented in \cite{mish93,heid92,elli92,heid94} are studied at 
finite temperature $T$ and 
nonzero baryon density $\rho$.
At nuclear matter saturation density $\rho_0$, the minimal model of 
\cite{mish93} exhibits two phases (the abnormal one at nearly vanishing nucleon mass and 
the normal phase at $m_{\ast} \simeq 0.7 m_N$),
which allows for a phase transition at high temperatures {\it or} high densities. 
The presence of vector mesons leads to an open phase boundary, and  the 
inclusion of dilatons makes the abnormal phase also 
mechanically stable.
 However, in the model abnormal solutions at $\rho_0$ exist only at 
unphysically high values of the compressibility (K $\stackrel{>}{\sim} 1400$ MeV). 
Therefore, the abnormal phase should be eliminated  by either including 
a $\omega-\sigma$-coupling or by replacing the quartic 
self-interaction with a logarithmic term ($\delta \neq 0$). 
In this case, no chiral phase transition can be found since the nucleon effective 
mass as order parameter increases at high densities and temperatures. 
It remains a challenge to construct 
a reasonable chiral model for nuclear matter which allows for the 
study of phase transitions.
First calculations done in an extension of the model to SU$(3)$  
are encouraging \cite{papa96}.\\
\begin{acknowledgements}
The authors are grateful to J. Eisenberg, C. Greiner,  
I. Mishustin, and K. Sailer for numerous fruitful discussions. This work 
was supported by Gesellschaft f\"ur Schwerionenforschung (GSI), 
Deutsche Forschungsgemeinschaft (DFG) and Bundesministerium f\"ur 
Bildung und Forschung (BMBF).  
\end{acknowledgements}

\begin{figure}
\vspace{-2cm}
\psfig{file=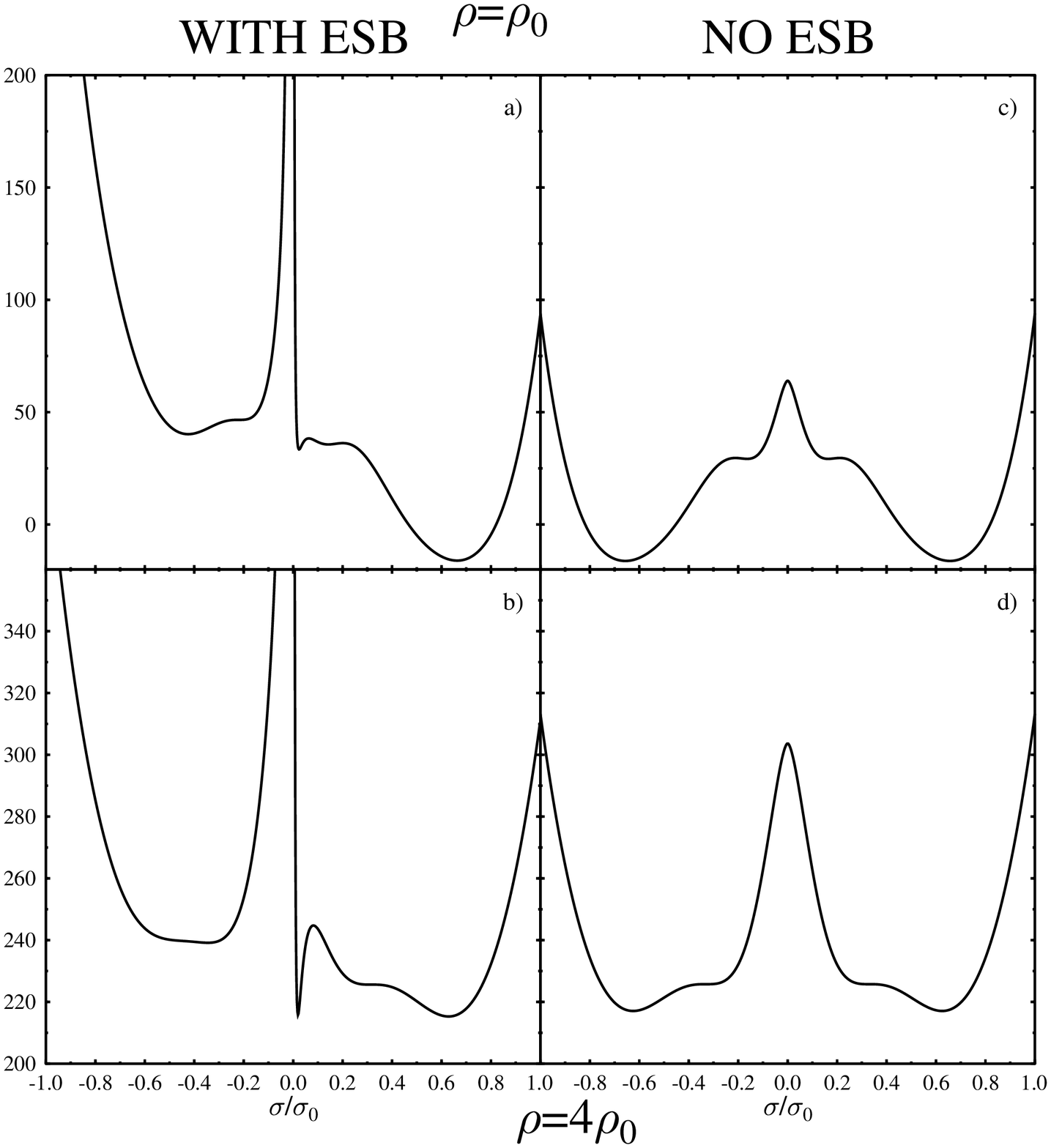,width=200mm}
\caption{\label{tall} Binding energy versus 
$\sigma/\sigma_0$ with (left) and without (right) 
explicit symmetry breaking (ESB) for saturation density $\rho_0$ 
(above) and $4 \rho_0$ (below) calculated with Set I}
\end{figure} 

\begin{figure}
\centerline{\psfig{file=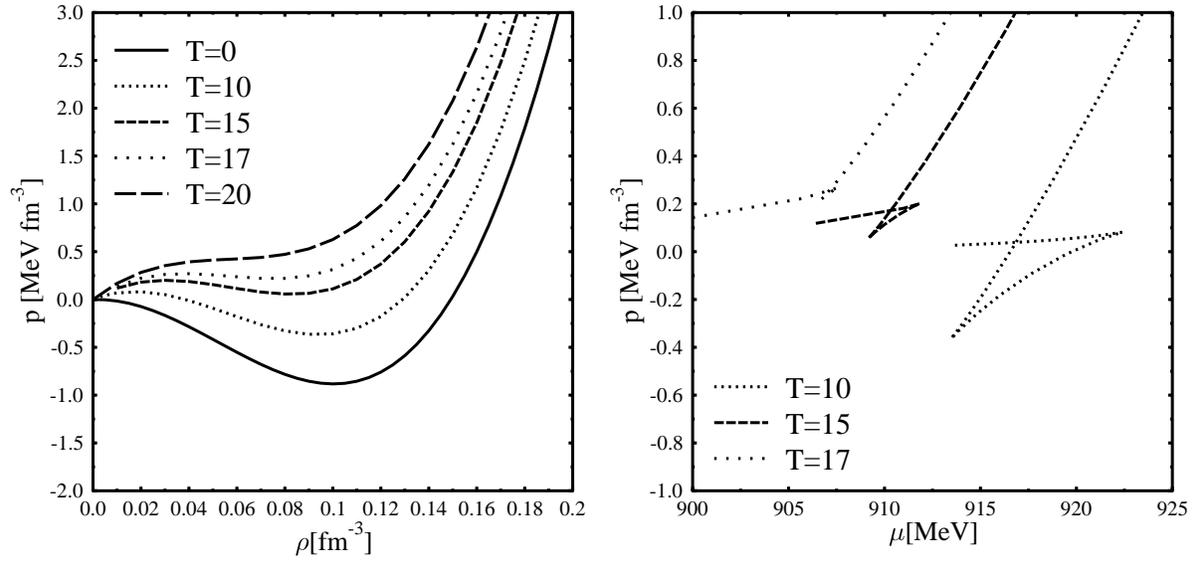,width=160mm}}
\caption
{\label{prho} Liquid-gas phase transition in the chiral $\sigma-\omega$ model
(calculated with Set V). }
\end{figure}

\begin{figure}
\centerline{\psfig{file=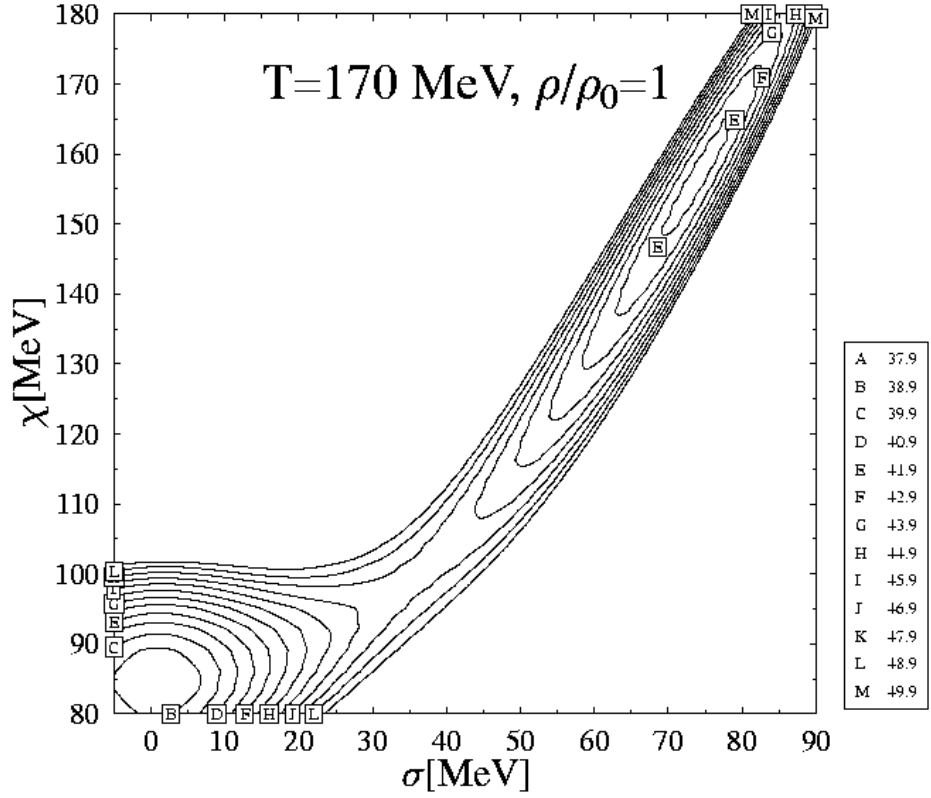,width=160mm}}
\par
\caption
{\label{min7} A contour plot of the free energy 
in the ($\chi, \sigma$)-plane. The abnormal and normal minimum are visible.}
\end{figure}

\begin{figure}
\vbox{
\centerline{\psfig{file=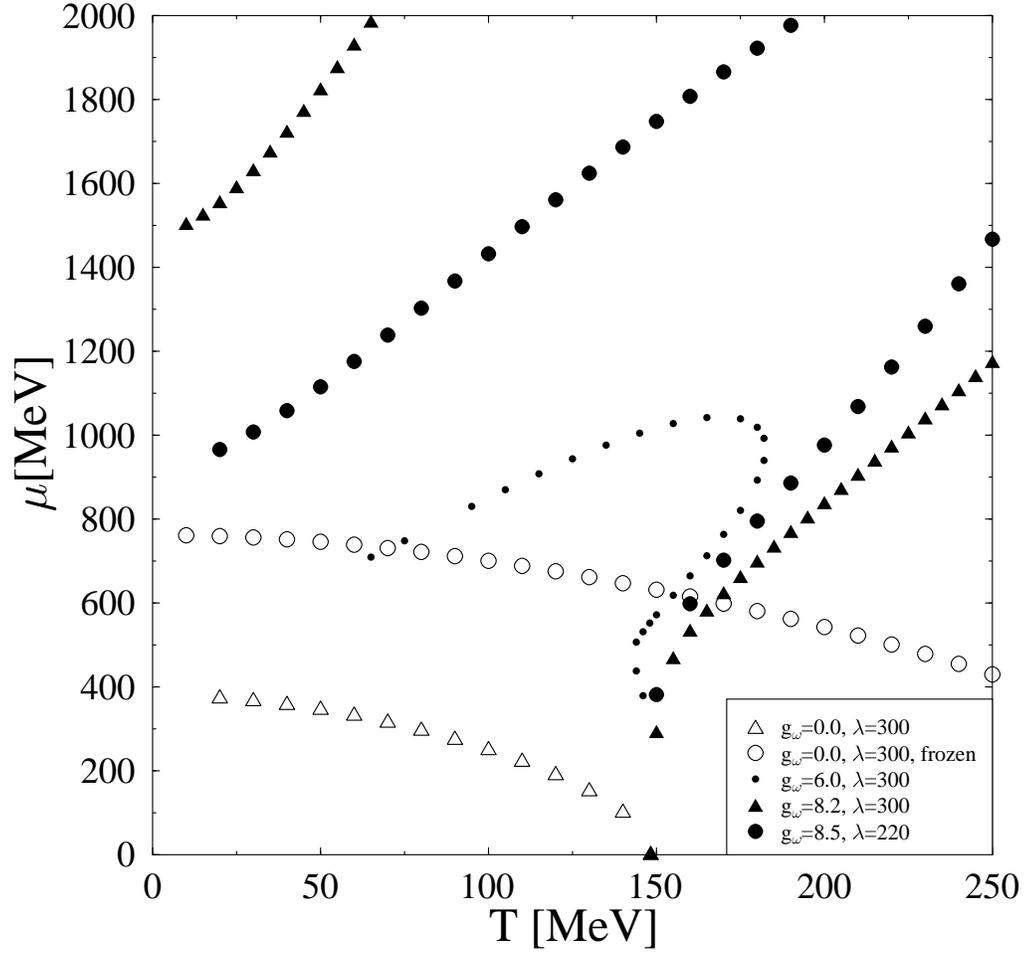,width=160mm}}}
\caption{\label{f300} Phase transition points for different values of the 
N-$\omega$ coupling constant in the ($\mu$, T)-plane}
\end{figure}

\begin{figure}
\vbox{
\centerline{\psfig{file=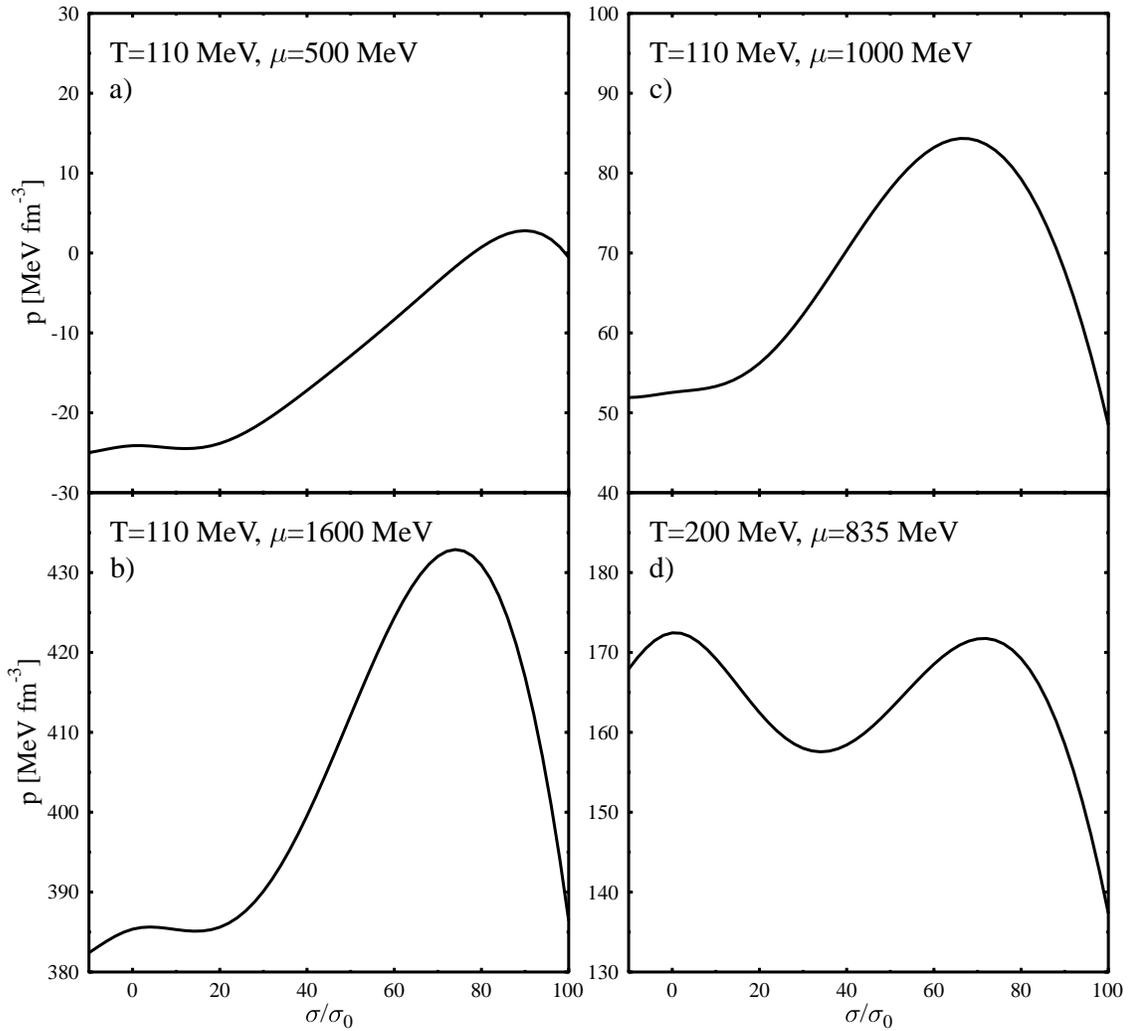,width=160mm}}}
\caption{\label{psigma} Pressure versus the scalar field $\sigma$. 
The abnormal solution 
does not exist far from the phase transition line. 
The barrier between normal and abnormal phase becomes well pronounced near the 
phase transition region.}
\end{figure}

\begin{figure}
\vspace{-3cm}
\hspace{-4cm}
\centerline{\psfig{file=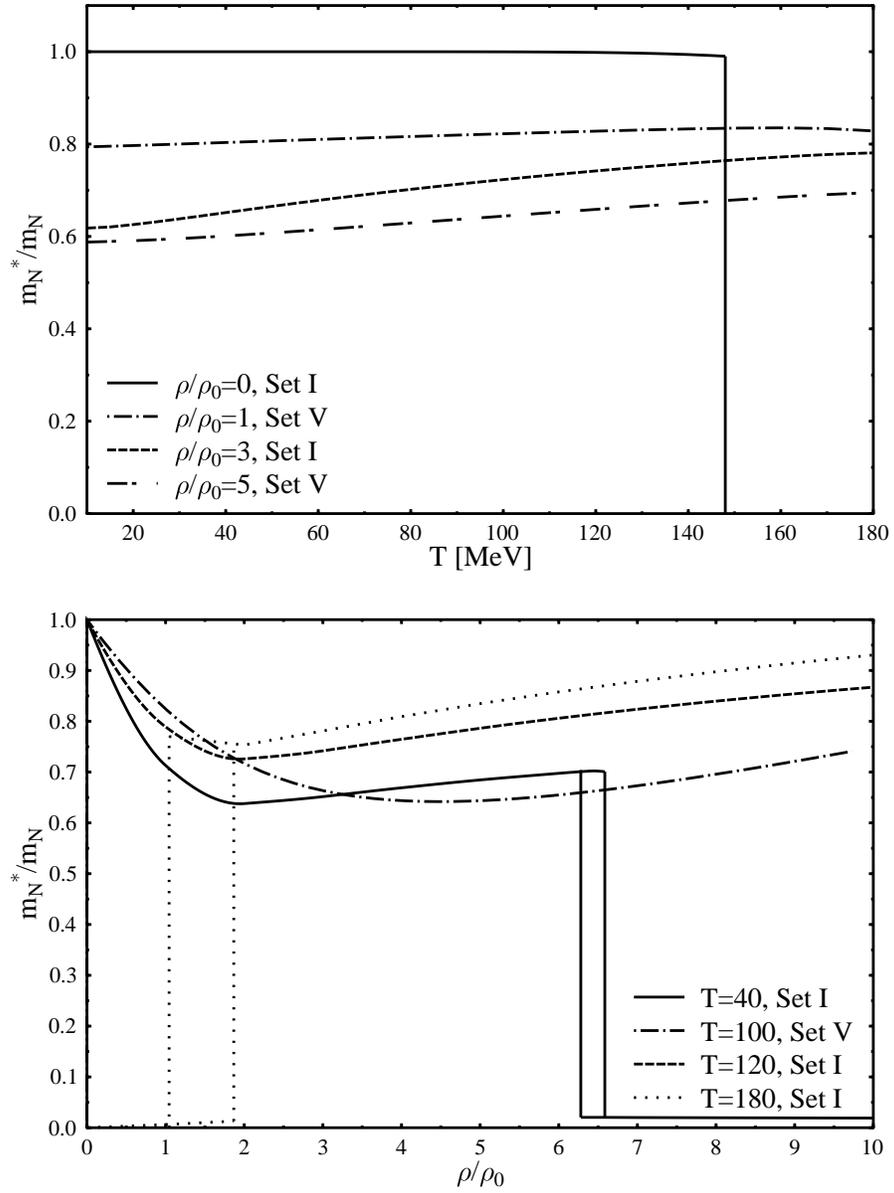,width=120mm}}
\caption {\label{emass} Effective nucleon mass versus temperature (above) and density (below)}
\end{figure}

\begin{figure}
\centerline{\psfig{file=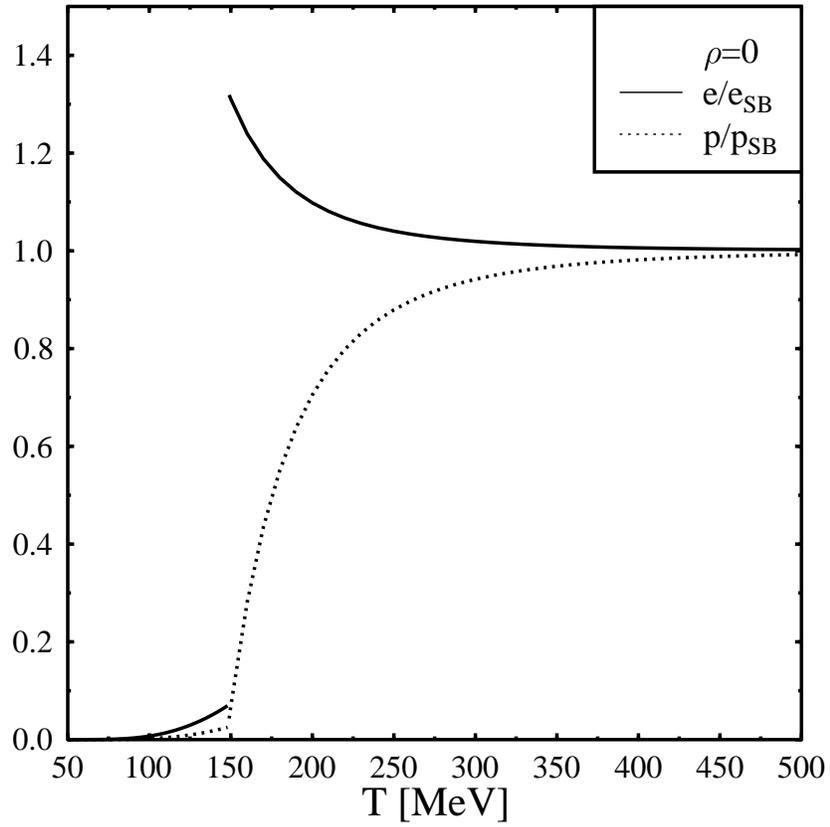,width=140mm,height=140mm}}
\caption
{\label{et4}High temperature limit of the energy density and pressure 
for zero density (Set I).}
\end{figure}

\begin{figure}
\centerline{\psfig{file=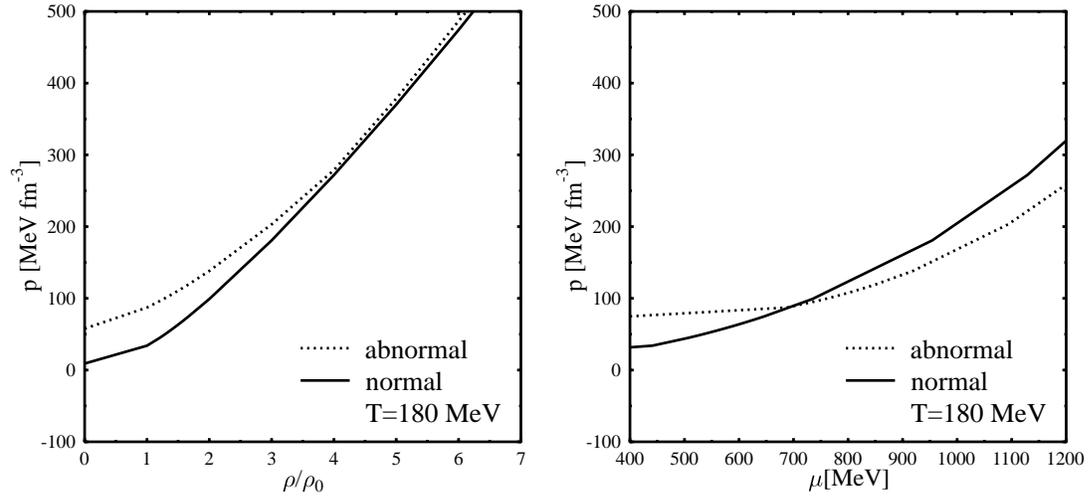,width=160mm }}
\caption{\label{prho2} Pressure as a function of density (left) and chemical 
potential (right), calculated with Set I. }
\end{figure}

\begin{figure}
\centerline{\psfig{file=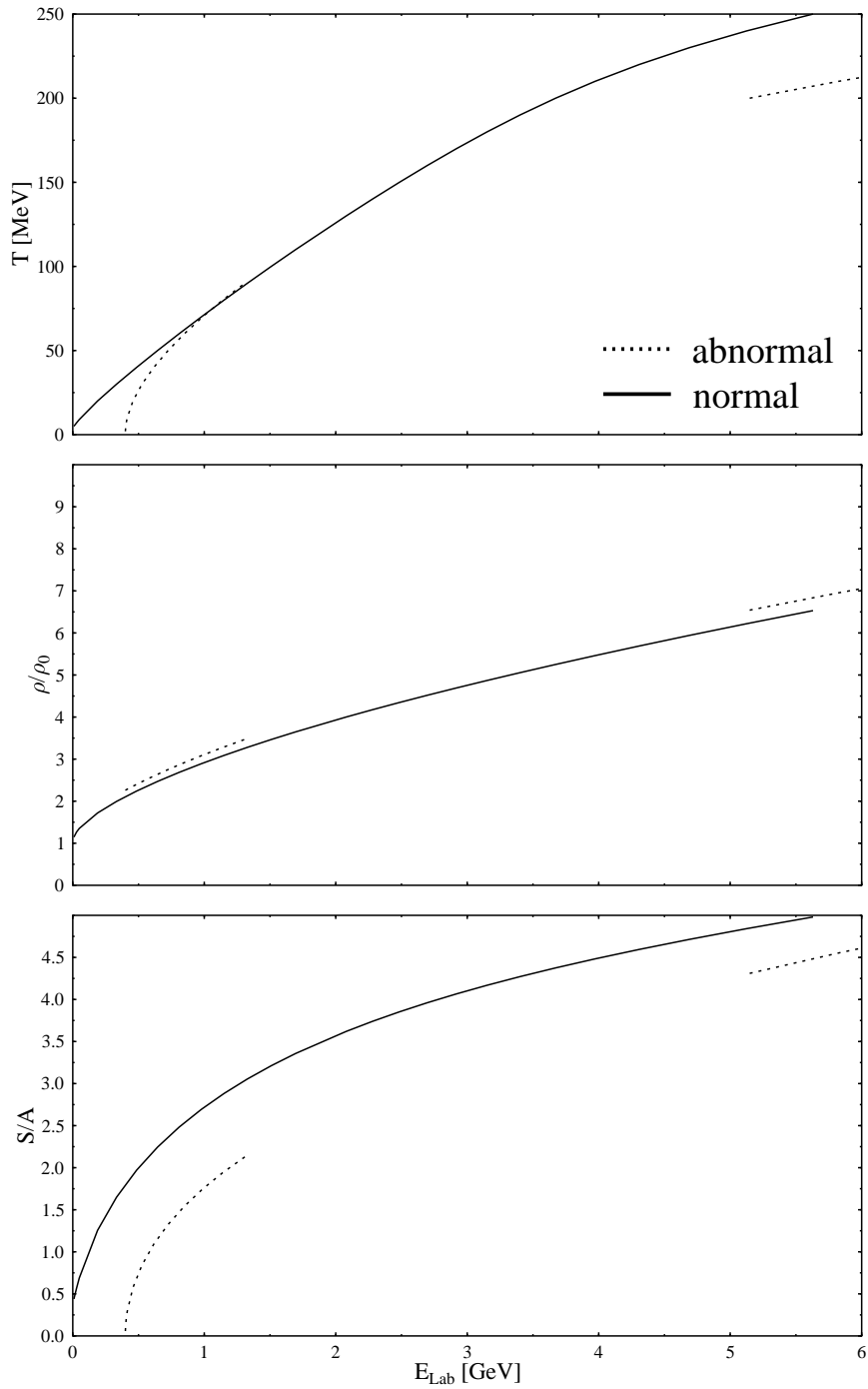}}
\caption{\label{taub3} Temperature, density and entropy per baryon as a function of the bombarding energy E$_{Lab}$ (Set I).} 
\end{figure}

\begin{figure}
\centerline{\psfig{file=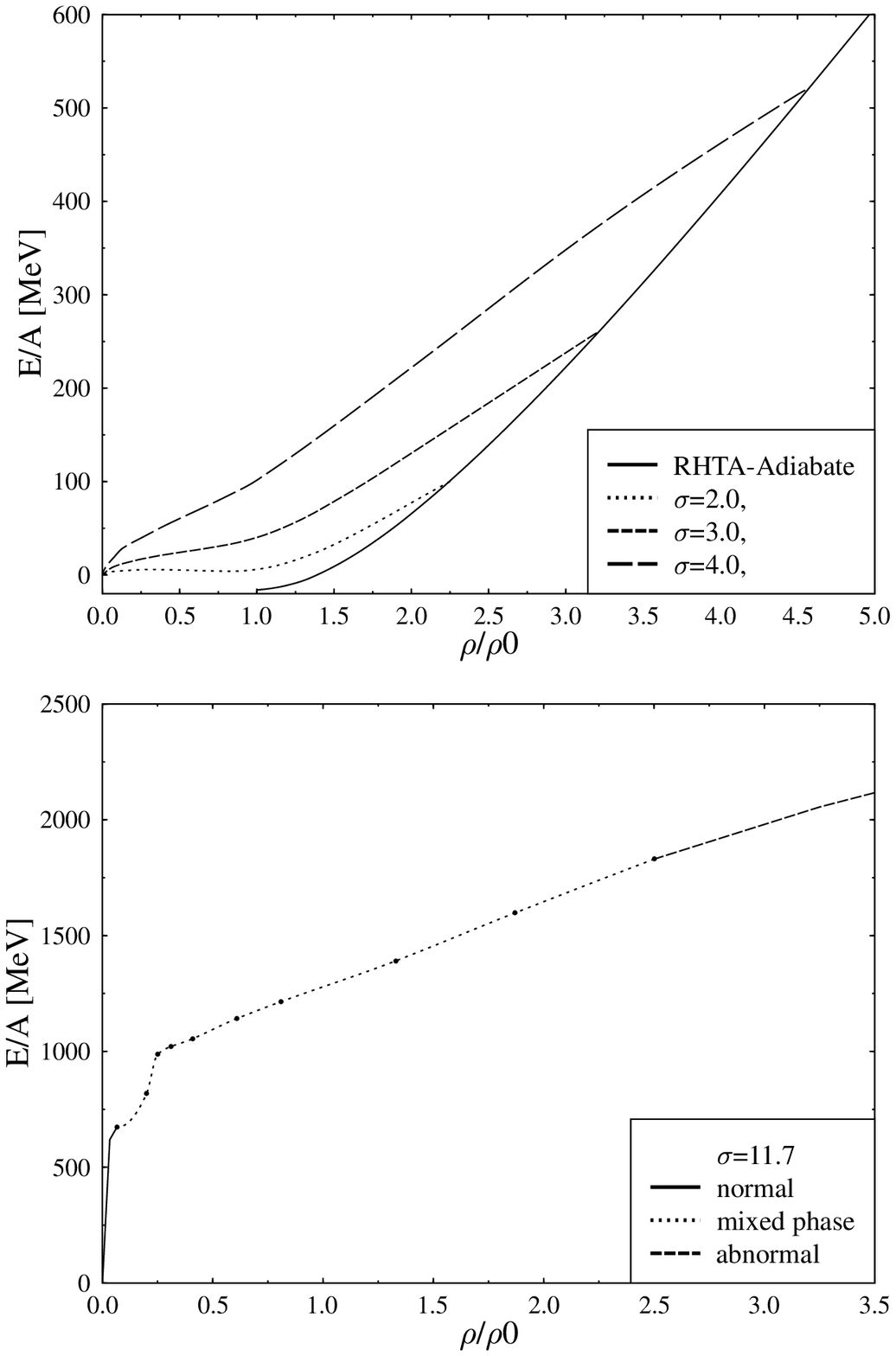}}
\caption{\label{expansion} Expansion of a compressed state along different isentropes starting from the normal phase  (a) and 
from the abnormal phase (b), respectively.}
\end{figure}

\begin{table}
\caption{\label{su2tab} Different parameter sets which describe nuclear 
matter  saturation} 
\begin{center}
\begin{tabular}{|c|c|c|c|c|c|c|c|c|c|} 
Set  &$\lambda$  & $\chi_0$ (MeV) & $g_{\omega}$
& $m_N^{\ast}/m_N$ & $\chi/\chi_0$& $K$ (MeV) & r & $33 \delta$ & $m_{\pi} (MeV)$\\ \hline
 I   & 300 & 189.3  & 8.2 & 0.66 & 0.71  & 1464  &  0  & 0 & 138  \\ 
 II  & 220 & 188.7  & 8.2 & 0.67 & 0.71  & 1403  & 0.5 & 0 & 138 \\
 III & 40  & 331.7  & 6.8 & 0.78 & 0.94  & 669   &  1  & 0 & 138  \\
 IV  & 0.84& 392.9  & 5.9 & 0.84 & 0.99  & 387   &  1  & 4 & 0   \\ 
 V   & 0   & 372.5  & 7.6 & 0.80 & 0.98  & 356   & 0.5 & 4 & 0   \\ 
 \end{tabular}
\end{center}
\end{table} 


\end{document}